\documentclass{ws-procs9x6}
\usepackage{amssymb}
\usepackage{amsfonts}

\begin{document}

\title{Shear Viscosity from the Effective Coupling of Gravitons}

\author{Rong-Gen Cai, Zhang-Yu Nie and Ya-Wen Sun}

\address{Institute of Theoretical Physics, Chinese Academy of Sciences,\\
Beijing 100190, China\\
E-mail: cairg,niezy,sunyw@itp.ac.cn}

\begin{abstract}
We review the progress in the holographic calculation of shear
viscosity for strongly coupled field theories. We focus on the
calculation of shear viscosity from the effective coupling of
transverse gravitons and present some explicit examples.
\end{abstract}


\bodymatter

\section{Introduction}\label{aba:sec1}
The AdS/CFT
correspondence\cite{Maldacena:1997re,Gubser:1998bc,Witten:1998qj,Aharony:1999ti}
provides a new way to solve strongly coupled field theory problems.
Especially, we can calculate the transport coefficients in the
hydrodynamic limit of the field theory from the gravity side. Shear
viscosity is one of these coefficients that have been vastly studied
through the holographic
method\cite{Policastro:2001yc,Buchel:2003tz,Kovtun:2004de,Kovtun:2003wp}.
In the early days, the shear viscosity was calculated in a lot of
theories dual to Einstein gravity with or without chemical
potential\cite{Maeda:2006by,Saremi:2006ep,Son:2006em,Mas:2006dy}. It
was found that the value of the shear viscosity over the entropy
density ratio $\eta/s$ is always $1/4\pi$. After considering large N
effects, this ratio gets positive corrections. Then it was
conjectured that there is a universal lower bound on the ratio
$\eta/s$ in nature(KSS bound\cite{Kovtun:2003wp}). Known experiments
such as the Quark-Gluon Plasma(QGP) in RHIC also satisfy this
conjecture. Then more attention has been payed to the study of the
universality of $\eta/s$ through holographic method, and the
$\eta/s$ was found to be still above $1/4\pi$ in
Ref.~\refcite{Buchel:2008ae,Buchel:2008sh,Myers:2008yi,
Buchel:2008wy,Buchel:2008ac,Buchel:2004di,Benincasa:2005qc,Cai:2008in}.
However, it was later found that the KSS bound can be slightly
violated in theories with gravity dual of more general theories of
gravity\cite{Brigante:2007nu}.

In the study of the $\eta/s$ through holographic method, it was
gradually found that the shear viscosity of the boundary field
theory only depends on the effective coupling of the transverse
gravitons valued on the horizon in the gravity
side\cite{Brustein:2007jj,
Brustein:2008cg,Brustein:2008xx,Iqbal:2008by,Cai:2008ph}. This
relation applies to systems dual to more general theories beyond
Einstein gravity, and may shed some light on the understanding of
the AdS/CFT correspondence in another point of view. We will shortly
review the progress of studies on this problem in this article, and
also introduce some interesting results using the language of the
effective coupling.

The plan of this paper is as follows. In the next section, we review
the progress in the study of shear viscosity using holographic
method. In Sec. 3, we introduce recent studies on the calculation of
shear viscosity from the effective coupling of transverse gravitons,
and give the result in AdS Gauss-Bonnet gravity. In Sec. 4 we
present some examples from the effective coupling.
\section{Shear viscosity from AdS/CFT}
The AdS/CFT
correspondence\cite{Maldacena:1997re,Gubser:1998bc,Witten:1998qj,Aharony:1999ti}
is a duality between the anti-de Sitter (AdS) spacetime and a
conformal field theory(CFT) living on the boundary of the AdS
spacetime. In details, suppose that we have a field $\phi$ moving in
the AdS spacetime and a corresponding operator $\mathcal {O}$ in the
CFT side, and the boundary value $\phi_0$ of $\phi$ coupled to the
boundary operator $\mathcal {O}$ as a source, then the partition
function of $\phi$ with boundary value $\phi_0$ in the gravity side
is equal to the generating functional of $\mathcal {O}$ with a
source $\phi_0$ in the CFT side
\begin{equation} \label{Dual}
Z(\phi_0)=\Big<e^{i\int \phi_0 \mathcal {O}}\Big>_{CFT}.
\end{equation}
In order to get some information in the CFT, we only need to study
the partition function $Z(\phi_0)$ in the gravity side. In the
AdS/CFT correspondence, the gravity system in the AdS side is
classical at low energies, so the partition function $Z(\phi_0)$ can
be just determined by the classical solution of the field $\phi$
\begin{equation}
Z(\phi_0)=e^{iS(\phi)},
\end{equation}
where $S(\phi)$ is the classical action of the classical solution
$\phi$ with boundary value $\phi_0$. Then with this correspondence,
we can easily get the correlation functions of the operator
$\mathcal {O}$ in the strongly coupled field theory by the standard
functional formulas.

In the point of view of the field theory, knowing the correlation
functions of operators, one can extract transport coefficients using
the Kubo formula. Applying this formula to the case of energy
momentum tensor, we can calculate the shear viscosity by
\cite{Policastro:2002se,Son:2007vk}
\begin{equation}\label{Kubo}
\eta=\lim_{\omega\rightarrow 0}\frac{1}{2\omega
i}\Big(G^A_{xy,xy}(\omega,0)-G^R_{xy,xy}(\omega,0)\Big),
\end{equation}
where $\eta$ is the shear viscosity, and the retarded Green's
function is defined by
\begin{equation}
G^R_{\mu\nu,\lambda\rho}(k)=-i\int d^4xe^{-ik\cdot x}\theta (t)
\langle[T_{\mu\nu}(x),T_{\lambda\rho}(0)] \rangle.
\end{equation}
The advanced Green's function can be related to the retarded Green's
function by
$G^A_{\mu\nu,\lambda\rho}(k)=G^R_{\mu\nu,\lambda\rho}(k)^{*}$.

Using the AdS/CFT duality and the Kubo formula, the shear viscosity
$\eta$ can be calculated from the gravity dual, and interesting
results have been obtained using this holographic method. It was
found that the ratio of shear viscosity over entropy $\eta/s$
calculated from gravity dual of Einstein gravity is always $1/4\pi$.
After considering some corrections from large N effect, the ratio
gets positive corrections. Notice that in all the known experiments,
the fluids measured all have a ratio $\eta/s$ larger than $1/4\pi$,
The authors of Ref.~\refcite{Kovtun:2003wp} conjectured that this
ratio may have a universal lower bound in real fluids in nature,
called the KSS bound.
\section{Shear viscosity and the effective coupling of transverse gravitons}
It has recently been conjectured that the shear viscosity depends on
the effective coupling of transverse gravitons valued on the
horizon\cite{Brustein:2008cg}. In Ref.~\refcite{Brustein:2007jj,
Brustein:2008cg,Brustein:2008xx} the authors conjectured that this
ratio is equal to a quotient of effective couplings of two different
polarizations of gravitons, $\kappa_{xy}$ and $\kappa_{rt}$ valued
on the horizon. Then the dependence of shear viscosity on the
effective coupling of transverse gravitons is confirmed in Ref.
~\refcite{Iqbal:2008by} using a membrane paradigm and in Ref.
~\refcite{Cai:2008ph} by calculating the on-shell action of the
transverse gravitons. In Sec. 3.1 we show the relation of the shear
viscosity and the effective coupling of transverse gravitons. In
Sec. 3.2 we discuss the case of Gauss-Bonnet gravity.
\subsection{Transverse Graviton}
For simplicity, we consider a five dimensional asymptotic AdS black
hole solution with a Ricci flat horizon to some kind of gravity
theory with or without matter fields. We write the metric as
\begin{equation}\label{black}
ds^2=-{g(u)(1-u)}dt^2+\frac{1}{h(u)(1-u)}du^2+\frac{r_h^2}{ul}(dx^2+dy^2+dz^2),
\end{equation}
where $r_h$ is the horizon radius and $u=r_h/r$. Then the horizon is
located at $u=1$ and the boundary lives at $u=0$. $g(u)$ and $h(u)$
are two functions regular at $u=1$ and $l$ is the AdS radius which
is related to the cosmological constant by $\Lambda =-6/l^2$.

The boundary of this AdS spacetime is a $\mathbb{R}^4$, on which a
CFT of some kind lives. To calculate the shear viscosity of this
CFT, we need the Green functions of the energy momentum tensor,
which we can get by making a small perturbation to the metric tensor
in the gravity side. For simplicity, we choose spatial coordinates
$x$, $y$, $z$, so that the momentum of the perturbations points
along the z-axis. Then the perturbations can be written as $h_{\mu
\nu}=h_{\mu \nu}(t, z, u)$. In this basis there are three groups of
gravity perturbations, each of which is decoupled from others: the
scalar, vector and tensor perturbations\cite{Kovtun:2005ev}. Here we
use the simplest one that can be used to calculate the shear
viscosity, the tensor perturbation $h_{xy}$. We use $\phi$ to denote
this perturbation $\phi=h^x_y$ and write $\phi$ in a basis as
$\phi(t,u,z)=\phi(u)e^{-i\omega t+ip z}$. Then by expanding the
gravity action to the second order with the background metric
(\ref{black}), one can get the effective action of the transverse
graviton
\begin{equation}\label{graviton}
S=\frac{1}{16\pi G}\int d^5x\sqrt{-g} K_{eff}(u)(\nabla_{\mu}
\phi\nabla^{\mu}\phi),
\end{equation}
where we assume the graviton to be a minimal coupled massless scalar
with an effective coupling $K_{eff}$. Then it can be proved that the
shear viscosity depends only on the effective coupling valued on the
horizon $K_{eff}(u=1)$ \cite{Cai:2008ph,Iqbal:2008by}.

In Ref.~\refcite{Iqbal:2008by}, the authors used the membrane
paradigm to analysis the universality of shear viscosity over
entropy ratio, and proved that the shear viscosity depends on the
coupling of the transverse gravitons valued on the horizon. This
 shows some relations between the fluid on the near horizon
slicing and the long wavelength low frequency limit of the boundary
CFT. The boundary theory can be seen as flowing from the horizon
physics throw some spacetime structures.

In the paper Ref.~\refcite{Cai:2008ph}, we directly calculated the
on-shell action of the transverse gravitons, and obtained the shear
viscosity expressed by the effective coupling of transverse
gravitons. In Einstein gravity, this effective coupling is always
$-1/2$, so the shear viscosity over entropy density is always
$1/4\pi$ in theories with Einstein gravity dual. But in modified
gravity theories, the effective coupling of transverse gravitons
could be different. For example, in gravity theory with $R^2$
corrections, the effective coupling of transverse gravitons gets
corrections, thus in the dual field theory, the universal low bound
$1/4\pi$ can be
violated\cite{Brigante:2007nu,Brigante:2008gz,Kats:2007mq,Neupane:2009sx,Neupane:2008dc}.
\subsection{KSS bound violation in Gauss-Bonnet gravity}
The $R^2$ corrections come from string theory or some other UV
completion of Einstein gravity. After field redefinitions these
theories reduce to Gauss-Bonnet gravity  up to the $R^2$ order. The
action of Gauss-Bonnet gravity is
\begin{equation}
S=\frac{1}{16\pi G}\int d^5 x\sqrt{-g}\Big(R-2\Lambda +\frac{\lambda
l^2}{2}R_{GB}\Big),
\end{equation}
in which
$$R_{GB}=(R^2-4R^{\mu\nu}R_{\mu\nu}+R^{\mu\nu\rho\sigma}R_{\mu\nu\rho\sigma})$$
is the Gauss-Bonnet term, $\lambda$ is the Gauss-Bonnet parameter
with a small positive value.

With the Gauss-Bonnet term, the action of the transverse gravitons
is still that of minimal coupled massless scalar, but the effective
coupling differs from the value in pure Einstein gravity(it is no
longer a constant and depends on the radial coordinate $u$). Thus
the ratio of shear viscosity over entropy density $\eta/s$ in the
dual field theory also differs from the universal value with
Einstein gravity dual. The ratio in the Gauss-Bonnet case is
calculated to be
\begin{equation}
\frac{\eta}{s}=\frac{1}{4\pi}(1-4\lambda).
\end{equation}
With a positive Gauss-Bonnet correction $\lambda>0$, the universal
lower bound of $1/4\pi$ will obviously be violated. But with
causality consideration from the CFT side, the Gauss-Bonnet
parameter $\lambda$ is constrained as
$\lambda<0.09$\cite{Brigante:2008gz}, and then the ratio $\eta/s$
will have a new lower bound as
\begin{equation}
\frac{\eta}{s}\geq \frac{4}{25\pi}.
\end{equation}
The calculation of $\eta/s$ from the effective coupling of the
transverse gravitons is convenient, and we will give more examples
of studying the ratio $\eta/s$ with this method. More discussions on
the shear viscosity from the effective coupling of transverse
gravitons could be found in
Ref.~\refcite{Pal:2009qg,Banerjee:2009ju,Pang:2009pd,Banerjee:2009wg,Banerjee:2009fm,Pang:2009ky,Paulos:2009yk}.

\section{More examples from effective coupling}

Since the violation of the KSS bound was discovered, the
universality of the bound of $\eta/s$ needs more attention, it would
be valuable to examine wether the ratio will get more corrections in
 more general gravity duals.

In our previous work\cite{Cai:2008ph,Cai:2009zv}, we calculated the
shear viscosity in the case of AdS Gauss-Bonnet gravity with $F^4$
term corrections of Maxwell field, and in AdS Gauss-Bonnet gravity
with dilaton coupling. In both the two cases, the ratio of shear
viscosity over entropy density gets more corrections from the AdS
Gauss-Bonnet case. We also considered the shear viscosity in the
extremal case in a recent work\cite{Cai:2009zn}. We give our results
here.
\subsection{Gauss-Bonnet Gravity with $F^4$ term
corrections of Maxwell field} In Ref.~\refcite{Ge:2008ni}, when
Maxwell field is added to the Gauss-Bonnet gravity, the ratio
$\eta/s$ gets positive corrections to the pure AdS Gauss-Bonnet
gravity case. Then in our work\cite{Cai:2008ph}, we further studied
the effect of non-linear term of Maxwell field on the shear
viscosity in the setup of Gauss-Bonnet gravity dual. We found that
the $F^4$ term has some effects on the ratio $\eta/s$. We further
found that with Maxwell field, with or without $F^4$ terms, the
ratio $\eta/s$ depends on the temperature $T$ in the way
\begin{equation}
\frac{\eta}{s}=\frac{1}{4\pi}(1-\frac{4\lambda\pi l^2}{r_h}T).
\end{equation}
This is very interesting, for that the effect of the Maxwell field
on $\eta/s$ is simply relating this ratio to the temperature of the
black hole.

\subsection{AdS Gauss-Bonnet gravity with dilaton coupling}

The dilaton field coupled to the Gauss-Bonnet gravity has
non-trivial contributions, for that the pure Gauss-Bonnet black hole
solution without the dilaton is not a solution of the case with
dilaton by simply requiring the dilaton being a constant. So it
would be interesting to see whether the dilaton field has
contributions to the shear viscosity in the dual field theory. In
Ref.~\refcite{Cai:2009zv}, we found that in the case of Gauss-Bonnet
gravity with dilaton coupling the effective action of the transverse
gravitons is still the type of minimally coupled massless scalar, so
the shear viscosity depends on the effective coupling of the
transverse gravitons. We calculated the ratio of shear viscosity
over entropy density, and the result is
\begin{equation}
\frac{\eta}{s}=\frac{1}{4\pi}\bigg(1-4\lambda
e^{(\tau-\gamma)\phi_d(1)}\Big(1-24\gamma^2\lambda
e^{(\tau-\gamma)\phi_d(1)}+3\gamma\tau\Big)\bigg).
\end{equation}
After considering the causality restriction, we can see that with
dilaton coupling, the new lower bound in AdS Gauss-Bonnet case of
the $\eta/s$ can be further violated in the parameter space of the
black hole solution.
\subsection{Shear viscosity at zero temperature}
It is well known that the extremal black hole is very different from
the ordinary black holes with a finite temperature. In a recent
paper\cite{Edalati:2009bi}, shear viscosity and other transport
coefficients of field theories with nonzero chemical potentials were
calculated at zero temperature in the background of extremal AdS RN
black holes using similar methods as in
Ref.~\refcite{Faulkner:2009wj} where properties of non-Fermi
fluid\cite{Rey:2008zz,Lee:2008xf,Liu:2009dm} are discussed  from
AdS/CFT, there the IR physics plays an important role in deriving
the UV physics on the boundary. In the extremal case of Einstein
gravity, the shear viscosity over entropy density is still $1/4\pi$.

We calculated the shear viscosity in field theories dual to
Gauss-Bonnet gravity at zero temperature with non-zero chemical
potential. We  confirmed that in the extremal case, the shear
viscosity also depends on the effective coupling of transverse
gravitons valued on horizon. Our results show that, the shear
viscosity over entropy density in field theory dual to Gauss-Bonnet
gravity at zero temperature is still $1/4\pi$, the same as that in
Einstein gravity.

\section{Conclusion}
In this paper we gave a brief review on the progress of calculating
the shear viscosity in strongly coupled field theory through
holographic method, with special attention to the connection between
the shear viscosity and the effective coupling of transverse
gravitons in the dual gravity side. This connection is applicable
for a wide class of gravity theories. We presented some examples of
this connection in Gauss-Bonnet gravity with Maxwell field, and the
case with dilaton coupling. We also discussed results with extremal
black holes.

\section*{Acknowledgments}

We would like to thank N. Ohta and Yan Liu for useful discussions
and collaborations in some relevant studies. RGC thanks the
organizers ICGA9 for a warm hospitality during the conference. This
work was supported partially by grants from NSFC, China (No.
10535060, No. 10821504 and No. 10975168) and a grant from MSTC,
China (No. 2010CB833004).

\bibliographystyle{ws-procs9x6}

\end{document}